\documentclass[aps,prb,twocolumn,groupedaddress]{revtex4-1}
\pdfoutput=1

\usepackage{amsmath}
\usepackage{amssymb}
\usepackage{bm}
\usepackage{graphicx}

\usepackage{hyperref}
\hypersetup{%
  breaklinks = {true},
  citecolor = {blue},
  colorlinks = {true},
  linkcolor = {red},
  pdfauthor = {\textcopyright\ Johan Nilsson},
  pdfcreator = {\LaTeX\ and \flqq hyperref\frqq},
  pdffitwindow = {true},
  pdfmenubar = {true},
  pdfpagelayout = {SinglePage},
  pdfstartview = {Fit},
  pdftoolbar = {true},
  plainpages = {false},
}

\newcommand{\la}{\langle}
\newcommand{\ra}{\rangle}
\newcommand{\ua}{\uparrow}
\newcommand{\da}{\downarrow}

\usepackage{color}
\usepackage{color}

\begin{document}


\title{Majorana fermion description of the Kondo lattice: \\ Variational and Path integral approach}

\author{Johan Nilsson \& Matteo Bazzanella}
\address{Department of Physics, University of Gothenburg, 412 96 Gothenburg,  Sweden}

\date{July, 2013}

\begin{abstract}

All models of interacting electrons and spins can be reformulated as theories of interacting Majorana fermions. We consider the Kondo lattice model that admits a symmetric representation in terms of Majorana fermions. In the first part of this work we study two variational states, which are natural in the Majorana formulation. At weak coupling a state in which three Majorana fermions tend to propagate together as bound objects is favored, while for strong coupling a better description is obtained by having deconfined Majorana fermions. This way of looking at the Kondo lattice offers an alternative phenomenological description of this model.
In the second part of the paper we provide a detailed derivation of the discretized path integral formulation of any Majorana fermion theory. This general formulation will be useful as a starting point for further studies, such as Quantum Monte Carlo, perturbative expansions, and Renormalization Group analysis. As an example we use this path integral formalism to formulate a finite temperature variational calculation, which generalizes the ground state variational calculation of the first part. This calculation shows how the formation of three-body bound states of Majorana fermions can be handled in the path integral formalism.

\end{abstract}

\maketitle

\section{Introduction}
\label{sec:intro}

One of the most studied models in condensed matter physics is the Kondo lattice model, which consists of a lattice of localized magnetic moments ($f$-spins) interacting with a sea of conduction electrons ($c$-electrons), via an exchange coupling. This situation is characteristic of many varieties of rare earth and actinide compounds, and it is believed to be responsible for the incredibly rich physics of these systems. The properties of this model system have been studied intensively both analytically and numerically, deploying many different techniques, see e.g. the reviews in Refs.~\onlinecite{Sigrist_RMP_1997,Gulacsi2004,Newns_Read_1987,Hewson_Book}.
The main issue encountered by any analysis of the Kondo lattice model is the difference between the algebras of the $f$-spins and the $c$-electrons. Because of this difference it is very hard to describe the two subsystems on equal footing.

In a recent work  we introduced a faithful fermionic representation of the Kondo lattice,\cite{nilsson2011a} that was based on the representation of the Hamiltonian in terms of Majorana fermion degrees of freedom (Majoranas). Using this representation it is possible to treat spins and electrons in a more symmetric way, without focusing on one subsystem in particular. The common (Clifford) algebra fulfilled by the Majorana fermions allows for the generation of electron-spin canonical transformations, which are otherwise quite involved when written in terms of the standard operators. 
Majorana fermion representations of spins is an old technique,\cite{Casalbuoni1976b,BerezinMarinov1977} that was recently popularized by its application in the influential Kitaev model.\cite{Kitaev2006} In the context of our study this representation has previously been used to study the conventional Kondo lattice,\cite{Viera1981} and was later used in works focusing on non-Fermi liquid behavior in modified Kondo impurity problems and lattice systems.\cite{Miranda1994,Coleman1995,Bulla1997} The fact that this representation can be used to study spin models and generate spin liquid states has also been known for quite some time.\cite{sacramento1988helmholtz,Tsvelik1992,Shastry_Sen_1997,Tsvelik_book,Coleman2003} To the best of our knowledge the treatment in Ref.~\onlinecite{nilsson2011a} is the first application of Majoranas in the standard Kondo lattice model that faithfully represents the model without the introduction of additional states or any need for constraints.

In this article we are going to improve on our previous results in two ways. In a first stage we will introduce a canonical transformation among the Majorana fermions, that will allow us to obtain better results for the upper bound on the ground state energy. To do this we consider the limits of weak and strong coupling. We will work under the assumption that at weak coupling the three Majorana fermions coming from the $f$-spins, stick together to form a coherent Majorana fermion, that becomes (to a good approximation) the appropriate degree of freedom relevant in this limit. This interpretation is naturally suggested by the form of the Hamiltonian. Increasing the coupling strength, the coherence among the three Majoranas decreases, and hence it becomes more natural to think of them as independent particles. We indicate this process with the term {\it deconfinement} of the Majorana fermions. Both these descriptions are exact in the extreme limits of zero coupling and no hopping (atomic limit).
In the second part of the paper we provide a detailed derivation of the imaginary time path integral formulation for a general Majorana fermion Hamiltonian. This will permit us to study the Kondo lattice model at non-zero temperatures and to take into account entropic effects. To do this
we will discuss in detail how to handle the three-body Majorana bound states, taking into account the effects due to the deconfinement process at a mean field level.

The paper is organized as follows: in section \ref{sec:canonicaltransformation} we briefly review the Majorana representation of the Kondo lattice; in section \ref{sec:operatorvariation} we will then immediately introduce the canonical transformation among the Majorana fermions and use it to find the best variational guess for the ground state. We also discuss the physical meaning of the trial states that we consider.
In section \ref{sec:MajoranaPathIntegral} we study the path integral formulation of a general theory expressed in terms of Majorana fermions and in section \ref{sec:trialdensitymatrix} we will
make use of this formalism to study our Kondo lattice Hamiltonian. At the end, in section \ref{sec:conclusions}, we provide a brief outlook and a summary of the results.

\section{Majorana fermion representation of the Kondo lattice model}
\label{sec:canonicaltransformation}

The Hamiltonian for the Kondo lattice, that we are going to discuss in this paper, is conventionally written as \cite{Sigrist_RMP_1997}
\begin{eqnarray}
\label{HamiltonianKLM}
\hat{H}_{KLM} = &-&t\sum_\sigma \sum_{\langle i,j \rangle}c^\dagger_{c,\sigma}(\mathbf{r}_i)c_{c,\sigma}(\mathbf{r}_j)+\mbox{h.c.}
\\
&-& \mu\sum_i\left[n_c(\mathbf{r}_i)-1\right]+J\sum_i\mathbf S_c(\mathbf{r}_i)\cdot \mathbf S_f(\mathbf{r}_i).\nonumber
\end{eqnarray}
The first line describes the hopping of the conduction electrons on the lattice, and $\sigma=\uparrow,\downarrow$ is the spin label. In the
second line the last term represents the (on-site) interaction between the local magnetic moment and the spin of the conduction electron, while the first term
is the chemical potential term for the conduction electrons, so that $n_c(\mathbf r_i)=\sum_\sigma c^\dagger_{c,\sigma}(\mathbf r_i)c_{c,\sigma}(\mathbf r_i)$ is the number operator. In the following we will consider only the \emph{half-filled} Kondo lattice model, that implies a value of zero for the chemical potential. The operator $\mathbf S_f(\mathbf r_i)$ represents the spin-$\frac{1}{2}$ object coming from the magnetic moment on site $\mathbf r_i$, that satisfies the usual SU(2) algebra $\left[S^a(\mathbf r_i),S^b(\mathbf r_j)\right]=i \delta_{ij} \epsilon^{abc}S^c(\mathbf r_i)$ and $\mathbf S^2=3/4$.\cite{Sakurai_QM} Here, and in the following,
we will work in units such that $\hbar=1$. The $c$-electron's spin operator is given by the standard representation
$\mathbf{S}_c(\mathbf r_i)=\frac{1}{2}\sum_{\sigma,\sigma^\prime} c^\dagger_{c,\sigma}(\mathbf{r}_i)\bm{\tau}_{\sigma,\sigma^\prime}c_{c,\sigma^\prime}(\mathbf{r}_i)$, with $\tau^a$ ($a=1,2,3$) the Pauli matrices.\cite{Sakurai_QM}

Following Ref.~\onlinecite{nilsson2011a} it is possible to represent the Hamiltonian (\ref{HamiltonianKLM}) in terms of Majorana fermions. In order to do that,
it is necessary to decompose the $c$-electron and the $f$-spins in terms of Majorana fermion degrees of freedom. It is well known (see, e.g. Refs.~\onlinecite{Coleman1995,Shastry_Sen_1997}), that to represent spin-$\frac{1}{2}$ operators, three species of Majorana fermions are needed: $\mu_a(\mathbf r_i)$, ($a=1,2,3$).
The Majorana fermions are real $\mu^\dagger_a=\mu_a$ and independent 
$\left\{\mu_a(\mathbf r_i),\mu_b(\mathbf r_j)\right\}=\delta_{ij} \delta_{ab}$.
It's straightforward to check that the operators
\begin{equation}
\label{eq:majornaspinrep1}
S_f^a (\mathbf{r}_i) = -i \epsilon^{abc} \mu_b (\mathbf{r}_i) \mu_c (\mathbf{r}_i)/2,
\end{equation}
satisfy the angular momentum algebra for spin-$\frac{1}{2}$ operators. It is fundamental to note that a fourth Majorana fermion can be formed as a non-linear combination of the three just introduced:
\begin{equation}
\label{eq:gamma0}
\gamma_0(\mathbf r_i)=2i\mu_1(\mathbf r_i)\mu_2(\mathbf r_i)\mu_3(\mathbf r_i).
\end{equation}
This is a proper Majorana fermion (with $\gamma_0^2=1/2$), that commutes with the $f$-spin operator $\left[\gamma_0(\mathbf r_i),\mathbf S_f(\mathbf r_j)\right]=0$. As in Ref.~\onlinecite{nilsson2011a}
we can now use this Majorana fermion, together with three other Majorana fermions, to represent the $c$-fermion operators:
\begin{equation}
\begin{split}
c_\ua^{\, }(\mathbf{r}_i) &= 
e^{i \frac{\bm{\pi} \cdot \mathbf{r}_i}{2} }
\frac{\gamma_1(\mathbf{r}_i) - i \gamma_2(\mathbf{r}_i)}{\sqrt{2}}
, \\ 
c_\da^{\, }(\mathbf{r}_i) &= 
e^{i \frac{\bm{\pi} \cdot \mathbf{r}_i}{2} }
\frac{-\gamma_3 (\mathbf{r}_i) - i \gamma_0 (\mathbf{r}_i)}{\sqrt{2}}.
\label{eq:cdeflattice}
\end{split}
\end{equation}
The difference from the usual Majorana representation of conventional fermion operators,\cite{Coleman1995,ColemanSchofield1995} lies in the fact that $\gamma_0(\mathbf{r}_i)$ is here the composite operator \eqref{eq:gamma0}. The exponential factors in \eqref{eq:cdeflattice} generate a site-dependent phase. This phase convention can be generalized to other bipartite lattices, it is here written for a simple cubic lattice and is useful to simplify the algebra and to emphasize the interpretation of $\gamma_0$ as the appropriate degree of freedom at small values of the coupling constant. After straightforward algebra, making use of \eqref{eq:majornaspinrep1}-\eqref{eq:cdeflattice} and assuming nearest-neighbor hopping on a simple cubic lattice, the Hamiltonian \eqref{HamiltonianKLM} can be rewritten as
\begin{equation}
\hat{H}_{KLM} = \hat{H}^{(2)} + \hat{H}^{(4)} + \hat{H}^{(6)},
\end{equation}
with
\begin{widetext}
\begin{eqnarray}
\hat{H}^{(2)} &=&
i t  \sum_{a=1}^3 \sum_{i} \sum_{d=1}^D \gamma_a (\mathbf{r}_i + \hat{\mathbf{x}}_d) \gamma_a (\mathbf{r}_i)
+  \frac{J}{4}  \sum_{a=1}^3 \sum_{i}  i \gamma_a (\mathbf{r}_i) \mu_a (\mathbf{r}_i) ,
\label{MajoranaHamiltonian2}\\ 
\hat{H}^{(4)} &=& -\frac{J}{2} \sum_i \bigl[ i \gamma_1 (\mathbf{r}_i) \mu_1 (\mathbf{r}_i) i \gamma_2 (\mathbf{r}_i) \mu_2 (\mathbf{r}_i) + \text{cyclic permutations } 1\rightarrow 2 \rightarrow 3 \rightarrow 1 \bigr], 
\label{MajoranaHamiltonian4}
\end{eqnarray}
where $\hat{\mathbf{x}}_d$ represents the lattice vectors (one for each dimension). A part in the kinetic term of the $c$-electrons is quadratic in $\gamma_0$'s, and it is therefore an operator of sixth order in $\mu_a$:
\begin{equation}\label{eq:H6}
\hat{H}^{(6)} 
= i t   \sum_{i} \sum_{d=1}^D \gamma_0 (\mathbf{r}_i + \hat{\mathbf{x}}_d) \gamma_0 (\mathbf{r}_i)
=
-i 4t   \sum_{i} \sum_{d=1}^D \mu_1 (\mathbf{r}_i + \hat{\mathbf{x}}_d)\mu_2 (\mathbf{r}_i + \hat{\mathbf{x}}_d)\mu_3(\mathbf{r}_i + \hat{\mathbf{x}}_d) \mu_1(\mathbf{r}_i)\mu_2(\mathbf{r}_i)\mu_3(\mathbf{r}_i).
\end{equation}
\end{widetext}
This formula is a direct consequence of the gauge choice \eqref{eq:cdeflattice} and points out the nature of $\gamma_0$ as a composite propagating object: the coherent nearest-neighbor hopping is the effect that dominates the physics of the $\mu_a$-Majoranas at weak coupling $J/t$. Therefore in this limit it is adequate to consider $\gamma_0$ and not the three independent $\mu_a$-Majoranas as the appropriate degree of freedom.
Vice-versa, in the strong coupling limit, the $\mu_a$-Majoranas shall be considered as independent objects that hybridize with the $\gamma_a$'s,
and the operator in \eqref{eq:H6} treated as a weak gluing interaction.
It is important to note that both these states do not break the spin rotational symmetry, that is present in the initial Kondo lattice Hamiltonian, while in Ref.~\onlinecite{nilsson2011a} the only good trial wave functions in the small $J/t$ limit that we considered were broken symmetry states. To improve our results, keeping spin rotational symmetry, we will consider the hybridization of $\gamma$ and $\mu$ degrees of freedom in both limits. To optimize the linear combination between the two species of Majoranas, we will make use of a variational mean-field approach in the rest of the paper, though the path integral formalism introduced in section \ref{sec:MajoranaPathIntegral} permits to go beyond it.

For future convenience, we now introduce the conventions that we will use to describe the system in momentum space. A generic Majorana fermion can be expressed in the momentum basis as
\begin{multline}
\gamma_a(\mathbf{r}_i) =
\frac{1}{\sqrt{N}} \sum_{\mathbf{k}}
e^{i \mathbf{k} \cdot \mathbf{r}_i} \gamma_a^{\,}(\mathbf{k})
\\
=  \frac{1}{\sqrt{N}} {\sum_{\mathbf{k}}}' \Bigl[
e^{i \mathbf{k} \cdot \mathbf{r}_i} \gamma_a^{\,}(\mathbf{k}) + e^{-i \mathbf{k} \cdot \mathbf{r}_i} \gamma_a^\dagger (\mathbf{k})
\Bigr]
.
\label{eq:fouriergamma}
\end{multline}
Here $N$ is the number of lattice sites and the prime on the sum denotes that only one of $\mathbf{k}$ and $-\mathbf{k}$ should be included in the sum. This convention brings to more readable formulas, because
the operators $\gamma(\mathbf{k})$ and $\gamma^\dagger(\mathbf{k})$ fulfill the algebra of standard fermionic operators $\{\gamma_a(\mathbf{k}),\gamma_b^\dagger(\mathbf{k}^\prime)\}= \delta_{\mathbf{k}\mathbf{k}'} \delta_{ab}$ and $\{\gamma_a(\mathbf{k}),\gamma_b (\mathbf{k}^\prime)\}=0$. However, it is often convenient to use the expression with the unprimed sum in the intermediate steps.
We will, for simplicity, consider lattices  with $N_d = 2M_d$ and $M_d$ odd in the following. Then the allowed values of $\mathbf{k}$ are
$\mathbf{k} = 2 \pi (\frac{n_1+1/2}{N_1} , \frac{n_2+1/2}{N_2} , \frac{n_3+1/2}{N_3} , \ldots) $.
This means that $\mathbf{k}$ and $-\mathbf{k}$ are distinct and that zero-modes are absent. The operators that appear in \eqref{MajoranaHamiltonian2} can readily be reformulated in $\mathbf{k}$-space, via a Fourier transform. The first term reads
\begin{multline}
\label{eq:HkinMajorana2}
i t  \sum_{a=1}^3 \sum_{i} \sum_{d=1}^D \gamma_a (\mathbf{r}_i + \hat{\mathbf{x}}_d) \gamma_a (\mathbf{r}_i) \\ =
\frac{3}{4} E_0 + \sum_{a=1}^3 {\sum_{\mathbf{k}}}' 
\epsilon_{\mathbf{k}}   \gamma_a^\dagger (\mathbf{k}) \gamma_a^{\,}(\mathbf{k}),
\end{multline}
were $E_0$ is the ground state energy for $J=0$ (in 1D $E_0 / N = -4 t / \pi$ in the thermodynamic limit $N \rightarrow \infty$) and
\begin{equation}
\epsilon_{\mathbf{k}}  = 2 t \sum_d \sin ( \hat{\mathbf{x}}_d \cdot \mathbf{k})
= 2 t \sum_d \sin ( k_d ).
\end{equation}
It is natural to choose the $\mathbf{k}$'s in the sum so that $\epsilon_{\mathbf{k}}  >0$ and hence the ground state has no $\gamma$-excitations.
For example in 1D this implies that $0<k<\pi$, while in 2D it corresponds to the square enclosed by the lines $-k_x < k_y < 2\pi - k_x$ and $-\pi + k_x < k_y < \pi + k_x$.
We will identify these subsets of the full Brillouin zone with the symbol BZ$'$. The other quadratic term in the Hamiltonian is the hybridization term, which also assumes a simple form in momentum space:
\begin{multline}
\frac{J}{4} \sum_{a=1}^3 \sum_i 
i \gamma_a (\mathbf{r}_i)  \mu_a (\mathbf{r}_i) 
\\
=
\frac{J}{4} \sum_{a=1}^3 {\sum_{\mathbf{k}}}' 
[
i \gamma_a^\dagger (\mathbf{k}) \mu_a^{\,}(\mathbf{k})
-i \mu_a^\dagger (\mathbf{k}) \gamma_a^{\,}(\mathbf{k})
].
\label{eq:H2hyb}
\end{multline}
The operators that appear in \eqref{MajoranaHamiltonian4} and \eqref{eq:H6} can also be expressed in $\mathbf{k}$-space. However, in this case it is more convenient to postpone their treatment until we have defined our mean-field decomposition scheme, to simplify these interaction terms as much as possible.

\section{Variational calculation}
\label{sec:operatorvariation}

To gain energy from the quadratic part of the Hamiltonian $\hat{H}^{(2)}$ [in particular from \eqref{eq:H2hyb}]  in the Majorana basis, it is natural to perform a rotation between $\gamma$'s and $\mu$'s. To do this we define rotated operators ($a = 1,2,3$) for each value of $\mathbf{k}$ in BZ$'$:
\begin{equation}
\begin{split}
\tilde{\mu}_a^{\,}(\mathbf{k}) &= 
\cos(\alpha_{\mathbf{k}}/2) \mu_a^{\,}(\mathbf{k}) + i \sin(\alpha_{\mathbf{k}}/2) \gamma_a^{\,}(\mathbf{k}) ,
\\
\tilde{\gamma}_a^{\,}(\mathbf{k}) &= 
\cos(\alpha_{\mathbf{k}}/2) \gamma_a^{\,}(\mathbf{k}) + i \sin(\alpha_{\mathbf{k}}/2) \mu_a^{\,}(\mathbf{k}) .
\label{eq:Majoranarotation1}
\end{split}
\end{equation}
Note that this implies that $\alpha_{-\mathbf{k}} = - \alpha_{\mathbf{k}}$ since we want to preserve standard anti-commutation relations. Applying the transformation to the quadratic piece we obtain
\begin{multline}
\hat{H}^{(2)} 
= \sum_{a=1}^3 {\sum_{\mathbf{k}}}' \bigr[ 
\tilde{\epsilon}_{\gamma}(\mathbf{k}) \tilde{\gamma}_a^{\dagger}(\mathbf{k}) \tilde{\gamma}_a^{\,}(\mathbf{k})
+
\tilde{\epsilon}_{\mu}(\mathbf{k}) \tilde{\mu}_a^{\dagger}(\mathbf{k}) \tilde{\mu}_a^{\,}(\mathbf{k})
\bigr]
\\
+
 \sum_{a=1}^3 {\sum_{\mathbf{k}}}' \tilde{V}(\mathbf{k})
[
i \tilde{\gamma}_a^\dagger (\mathbf{k}) \tilde{\mu}_a^{\,}(\mathbf{k})
-i \tilde{\mu}_a^\dagger (\mathbf{k}) \tilde{\gamma}_a^{\,}(\mathbf{k})
] 
+ \frac{3}{4} E_0 , 
\label{eq:HquadraticOriginal1}
\end{multline}
where for generic $\alpha_{\mathbf{k}}$ we have
\begin{eqnarray}
\tilde{\epsilon}_{\gamma}(\mathbf{k}) &=& 
\cos^2(\alpha_{\mathbf{k}}/2) \epsilon_{\mathbf{k}} + \sin(\alpha_{\mathbf{k}}) J/4, 
\nonumber \\
\tilde{\epsilon}_{\mu}(\mathbf{k}) &=& 
\sin^2(\alpha_{\mathbf{k}}/2) \epsilon_{\mathbf{k}} - \sin(\alpha_{\mathbf{k}}) J/4 , 
\\
\tilde{V}(\mathbf{k}) &=& 
\cos(\alpha_{\mathbf{k}}) J/4 -\sin(\alpha_{\mathbf{k}})  \epsilon_{\mathbf{k}} / 2 
.
\nonumber
\end{eqnarray}
Note that $\tilde{\epsilon}_{\mu}(\mathbf{k}) < 0$ when $0 < \tan(\alpha_{\mathbf{k}}/2) < J /(2  \epsilon_{\mathbf{k}})$.
The choice  $\tan(\alpha_{\mathbf{k}}) = J / (2 \epsilon_{\mathbf{k}})$ diagonalizes the quadratic part of the Hamiltonian, i.e., it sets $\tilde{V}(\mathbf{k}) = 0$. However, as we shall see, this is not a good choice for small $J/t$ since the cost in kinetic energy of the $\gamma_0$'s is then too large. It is also convenient to introduce real space Majorana operators corresponding to the $\tilde{\mu}(\mathbf{k})$'s and the $\tilde{\gamma}(\mathbf{k})$'s. To do this we use the same convention as in \eqref{eq:fouriergamma}, but with tildes on the operators. We also define rotated versions of the three-body  bound state and spins via 
\begin{equation}
\begin{split}
\tilde{\gamma}_0 (\mathbf{r}_i) &= 2i \tilde{\mu}_1 (\mathbf{r}_i) \tilde{\mu}_2 (\mathbf{r}_i) \tilde{\mu}_3 (\mathbf{r}_i),
\\
\tilde{S}_a (\mathbf{r}_i) &= -i \epsilon_{abc} \tilde{\mu}_b (\mathbf{r}_i) \tilde{\mu}_c (\mathbf{r}_i)/2.
\end{split}
\end{equation}
In the real space basis the relation between operators are ($a = 1,2,3$)
\begin{equation}
\begin{split}
\mu_a(\mathbf{r}_i) &= \sum_{j} 
\Bigl[
\mathcal{A}_{ij} \tilde{\mu}_a(\mathbf{r}_j) +  
\mathcal{B}_{ij} \tilde{\gamma}_a(\mathbf{r}_j)
\Bigr],
\\
\gamma_a(\mathbf{r}_i) &= \sum_{j} 
\Bigl[
\mathcal{A}_{ij} \tilde{\gamma}_a(\mathbf{r}_j) +  
\mathcal{B}_{ij} \tilde{\mu}_a(\mathbf{r}_j)
\Bigr],
\label{eq:Majoranarotation2}
\end{split}
\end{equation}
with matrix elements given by
\begin{equation}
\begin{split}
\mathcal{A}_{ij} &=
\frac{2}{N} {\sum_{\mathbf{k}}}' \cos(\alpha_{\mathbf{k}}/2) \cos (\mathbf{k} \cdot \mathbf{r}_{ij} ),
\\
\mathcal{B}_{ij} &=
\frac{2}{N} {\sum_{\mathbf{k}}}' \sin(\alpha_{\mathbf{k}}/2) \sin(\mathbf{k} \cdot \mathbf{r}_{ij} ).
\label{eq:transformationmatrices2}
\end{split}
\end{equation}
Since the Hamiltonian is invariant under inversion $\mathbf{k} \rightarrow \bm{\pi} - \mathbf{k}$ it is natural to assume that $\alpha_{\mathbf{k}} = \alpha_{\bm{\pi} - \mathbf{k}}$, which implies unbroken inversion symmetry. A consequence of this choice is that the matrix elements of $\mathcal{A}$ ($\mathcal{B}$) are nonzero only when the two indexes belong to the same (different) sublattice. This observation will simplify the algebra in the following.

\subsection{Confined trial state with three-body bound states and totally uncorrelated rotated spins}
\label{subsec:trial1}

We will consider two simple trial states for the ground state of the half-filled 1D Kondo lattice, although similar calculations are easily performed in higher dimensions as well. Both states  \emph{do not break spin rotational symmetry}. The first trial state, which works better for small $J/t$, has a definite fermion parity for each of the flavors $\tilde{\gamma}_a$ ($a = 0, 1, 2, 3$). This implies that all trial state averages involving an odd number of $\tilde{\gamma}_a$'s will vanish. The state we consider has a definite occupation of each fermion state and totally uncorrelated (rotated) spins, explicitly 
\begin{equation}
\begin{split}
\la n_{\tilde{\gamma}_a}(\mathbf{k}) \ra &= 0 \; {\rm or \; } 1,   \qquad {\rm for} \quad a = 0,1,2,3,
\\
\la \tilde{S}_a (\mathbf{r}_i ) \tilde{S}_b (\mathbf{r}_j ) \ra &= \delta_{ab} \delta_{ij}/4 .
\label{eq:localspincorrelations}
\end{split}
\end{equation}
The variational energy of this state is easily expressed in terms of the $\mathcal{A}$ and $\mathcal{B}$ matrices in real space and the angles $\alpha_{\mathbf{k}}$. The averages we need are
\begin{widetext}
\begin{eqnarray}
\la \hat{H}^{(2)} \ra &=&\sum_{a=1}^3 {\sum_{\mathbf{k}}}' 
\Bigl(
\tilde{\epsilon}_{\gamma}(\mathbf{k}) \la n_{\tilde{\gamma}_a}(\mathbf{k}) \ra
+
\tilde{\epsilon}_{\mu}(\mathbf{k}) /2
\Bigr)  + 3 E_0 /4,
 \nonumber  \\
\la \hat{H}^{(4)}\ra &=& -\frac{J}{2 N}
\Bigl(
{\sum_{\mathbf{k}}}'  \sin(\alpha_{\mathbf{k}})
\la n_{\tilde{\gamma}_1}(\mathbf{k}) - \frac{1}{2} \ra \Bigr)
\Bigl(
{\sum_{\mathbf{k}}}'  \sin(\alpha_{\mathbf{k}})
\la n_{\tilde{\gamma}_2}(\mathbf{k}) - \frac{1}{2} \ra \Bigr)  + 
\text{cyclic permutations } 1 \rightarrow 2 \rightarrow 3 \rightarrow 1, 
\label{eq:H6average1} \\
\la \hat{H}^{(6)} \ra &=& \frac{t}{N} \sum_{i,j,l,d} \mathcal{A}^3_{ij}  \mathcal{A}^3_{il}  
{\sum_{\mathbf{k}}}' 2 \sin \bigl( \mathbf{k} \cdot (\mathbf{r}_{jl} + \hat{\mathbf{r}}_d ) \bigr)
\la n_{\tilde{\gamma}_0}(\mathbf{k})- \frac{1}{2} \ra
 - 4 t N \sum_d \prod_{a=1}^3
\Bigl(
\frac{1}{N}{\sum_{\mathbf{k}}}'  2 \sin(k_d) \sin^2
(\frac{\alpha_{\mathbf{k}}}{2})
\la n_{\tilde{\gamma}_a}(\mathbf{k})- \frac{1}{2} \ra \Bigr).
\nonumber 
\end{eqnarray}
\end{widetext}
We now consider the state with zero fermion excitations $\la n_{\tilde{\gamma}_a}(\mathbf{k}) \ra = 0$, since it has the lowest energy. The resulting energy functional (measured with respect to the ground state at $J=0$) becomes
\begin{multline}
\frac{\Delta E_{var}[\alpha_{\mathbf{k}}]}{N} = \frac{3}{2 N}{\sum_{\mathbf{k}}}' \tilde{\epsilon}_{\mu}(\mathbf{k}) 
+  \frac{t}{\pi} (1-\mathcal{A}_{00}^6 )
\\
- \frac{3 J}{8} 
\Bigl( \frac{1}{N}{\sum_{\mathbf{k}}}'  \sin(\alpha_{\mathbf{k}})\Bigr)^2
+ \frac{4 t}{3 \pi} \mathcal{A}_{00}^3 \mathcal{A}_{02}^3 + \ldots 
\label{eq:variationalenergyGlued}
\end{multline}
In principle the full expression should be optimized with respect to $\alpha_{\mathbf{k}}$, but a detailed analysis shows that for small values of $J/t$ the result is dominated by the first two terms. Moreover, in this limit we find that $ \mathcal{A}_{ij}, \mathcal{B}_{ij},  \ll \mathcal{A}_{00} \approx 1$ for all $i \neq j$ . Minimizing just those two terms, using the fact that $\mathcal{A}_{00} \approx 1$, the optimal angles are to a good approximation given by the solution of
\begin{equation}
\frac{4 t}{\pi} \sin(\alpha_{\mathbf{k}}/2) + \frac{\epsilon_{\mathbf{k}}}{2} \sin(\alpha_{\mathbf{k}})
=  \frac{J}{4} \cos(\alpha_{\mathbf{k}}) .
\label{eq:variationalparametersmallJ}
\end{equation}
An even simpler trial state is to take a constant angle $\alpha_{\mathbf{k}} = \alpha$. In this case the total energy functional is
\begin{multline}
\frac{\Delta E_{var}}{N} = 3 \left( \frac{t}{\pi} \sin^2(\alpha/2) - \frac{J}{16} \sin(\alpha) \right)
- \frac{3 J \sin^2(\alpha)}{32} 
\\
+ \frac{t}{\pi} [ 1 -\cos^6(\alpha/2) ]
+ \frac{4 t}{\pi^3} \sin^6(\alpha/2),
\label{eq:variationalenergy_bs}
\end{multline}
and the optimization can be performed analytically. In  Fig~\ref{fig:variationalenergy} we plot the variational energy with numerically optimized $\alpha_{\mathbf{k}}$ for the full $\Delta E_{var} [\alpha_{\mathbf{k}}]$, which is slightly better than the restricted variational states corresponding to \eqref{eq:variationalparametersmallJ} and \eqref{eq:variationalenergy_bs}. It is also compared with a conventional antiferromagnetic state in which the $f$-spins are locked into a static N\'{e}el order, and the SO(2)-symmetric state of Ref.~\onlinecite{nilsson2011a}. Both these states break spin rotational symmetry, in contrast to the trial states that we focus on here. The variational energy  of the N\'{e}el ordered state is given by \cite{Fazekas_Muller_Hartmann1991}
\begin{equation}
\frac{\Delta E_{{\rm AF}}}{N} =  - \frac{2}{N} {\sum_{\mathbf{k}}}'  \Bigl( \sqrt{(J/4)^2 + \epsilon^2(\mathbf{k})} - \epsilon(\mathbf{k}) \Bigr) .
\end{equation}
Both the energy of the N\'{e}el state, the SO(2)-state, and the the energy of our trial state \eqref{eq:variationalenergyGlued} reproduce the correct leading constant term in the limit $J/t \rightarrow 0$. They all approach this value with corrections of order $J^2 / t$ to logarithmic accuracy. These corrections make the N\'{e}el and SO(2) states slightly lower in energy for small but finite $J/t$. The SO(2) state is also energetically favorable to the N\'{e}el state, especially for larger values of $J/t$ where it can gain more of the singlet energy. The main point we wish to make is that the confined state does not break spin-rotational invariance and is comparable in energy to the broken-symmetry states for small $J/t$, especially if we allow for non-trivial (rotated) spin correlations. Since spin-rotational invariance is unbroken in the 1D Kondo lattice,\cite{LossLeggett2011} the confined state might provide a better approximation to the actual situation than states with locally broken symmetry, like for example mean-field states with local order that becomes disordered on longer length scales due to fluctuations.

\subsection{Trial state with independent deconfined Majorana excitations}
\label{subsec:trial2}

\begin{figure}
\centering
\includegraphics[scale=.85]{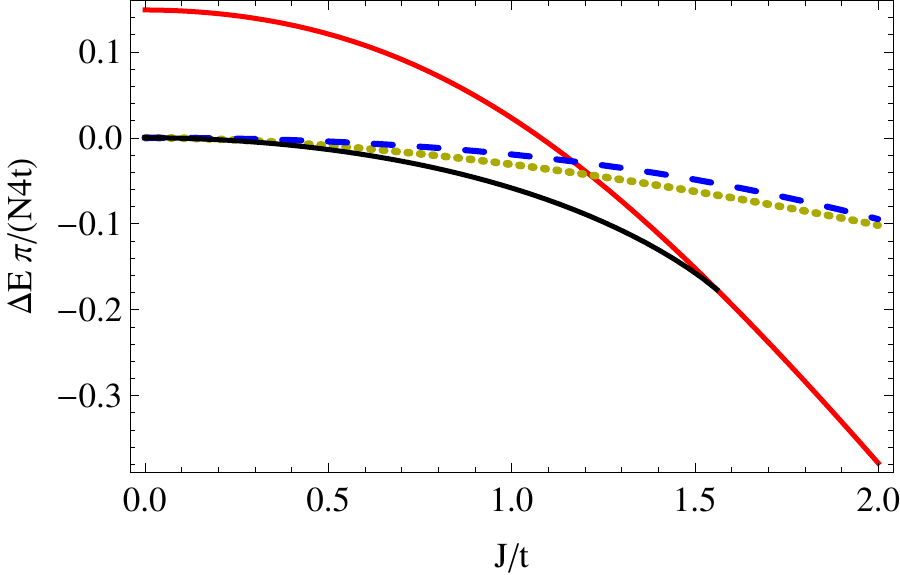}
\caption{
Variational energies for the half-filled 1D Kondo lattice for the different variational states.
Solid red line: deconfined Majorana state;
Dashed blue line: state with confined 3-body Majorana bound state;
Dotted yellow line: broken symmetry N\'{e}el ordered state.
Black solid line: broken symmetry SO(2)-state, taken from Ref.~\onlinecite{nilsson2011a}.
\label{fig:variationalenergy}
}
\end{figure}

For large values of $J/t$ it is important to gain energy from the dominating local term. This can be achieved by considering the trial state with
\begin{equation}
\begin{split}
\la n_{\tilde{\gamma}_a}(\mathbf{k}) \ra &= 0  \qquad {\rm for} \quad a = 1,2,3,
\\
\la n_{\tilde{\mu}_a}(\mathbf{k}) \ra &= 1   \qquad {\rm for} \quad a = 1,2,3.
\end{split}
\end{equation}
The calculation of the trial energy functional of this state is straightforward, with the result
\begin{multline}
\frac{\Delta E_{var}[\alpha_{\mathbf{k}}]}{N} =
\frac{3}{N} {\sum_{\mathbf{k}}}' \tilde{\epsilon}_{\mu}(\mathbf{k}) 
- \frac{3 J}{2} \Bigl( \frac{1}{N}{\sum_{\mathbf{k}}}'  \sin(\alpha_{\mathbf{k}})\Bigr)^2 - \frac{E_0}{4 N}
\\
-4 t \sum_d \Bigl( \frac{1}{N}{\sum_{\mathbf{k}}}'  \sin(k_d) \cos(\alpha_{\mathbf{k}})\Bigr)^3 .
\end{multline}
This is the O(3) state introduced in Ref.~\onlinecite{nilsson2011a} in a different notation. It is also displayed in Fig.~\ref{fig:variationalenergy} and is clearly favored with respect to the other considered states for large values of $J/t$. In fact it gives the leading term correctly in the limit $J/t \rightarrow \infty$.

\subsection{Trial states with non-trivial spin correlations}
\label{sec:variationspincorrelation}

The trial state of section~\ref{subsec:trial1} can be further improved by allowing for non-trivial spin correlations. Indeed, if we consider a state where the fermion subsystem is in the ground state, but leaving the state of the (rotated) spin system undetermined, the expectation value of the Hamiltonian with respect to the fermions will generate a spin Hamiltonian that depends on the variational parameters $\alpha_{\mathbf{k}}$. The best trial state is then obtained by setting the spin state to the ground state of this spin Hamiltonian. Optimizing the resulting trial energy with respect to $\alpha_{\mathbf{k}}$ the best possible trial state in this class is obtained. We leave this calculation for a later study since finding the ground state of the spin Hamiltonian is a highly non-trivial problem. We have checked that antiferromagnetic spin-spin correlations will improve the variational energy though.

\subsection{Physical characterization}

\begin{figure}
\centering
\includegraphics[scale=.65]{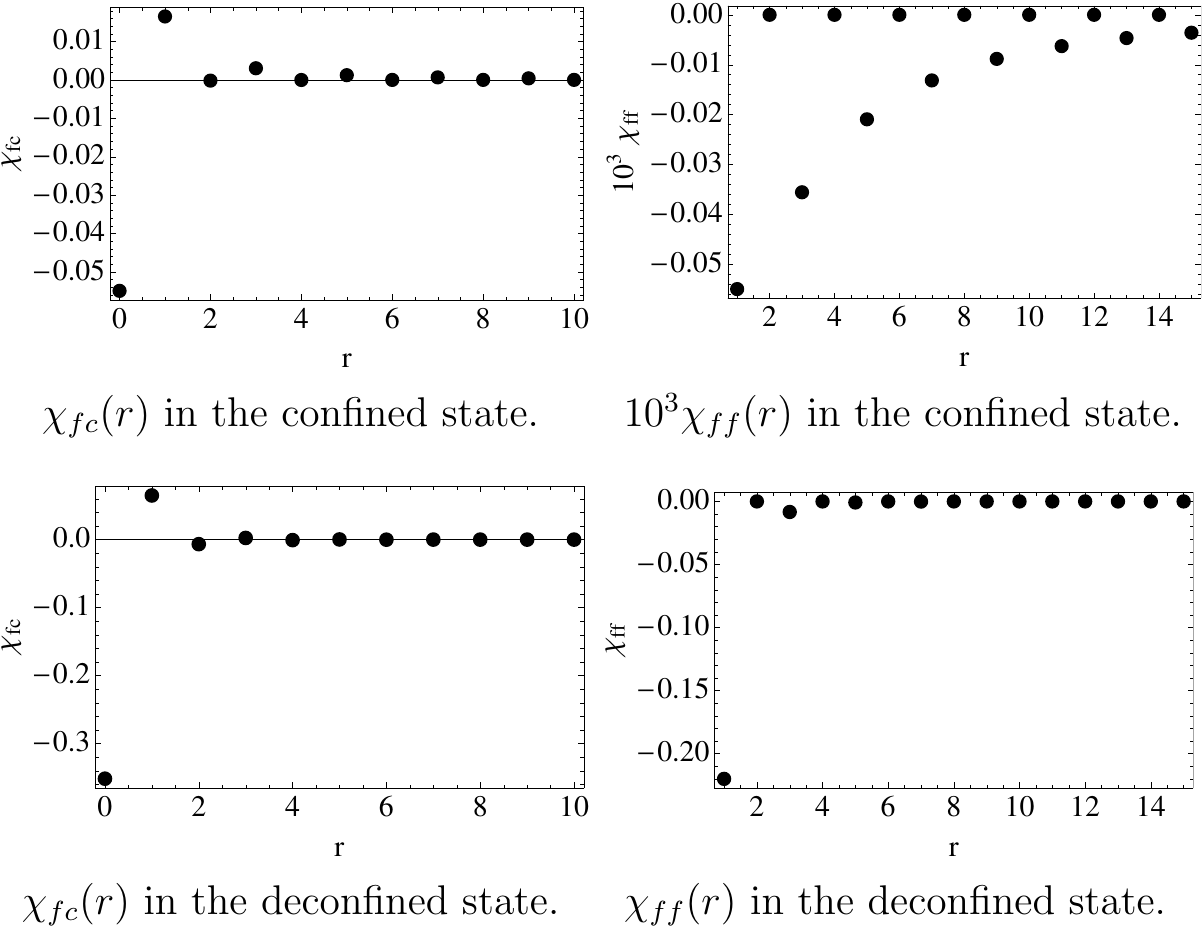}
\caption{
Spin-spin correlation functions for $J=t$ in the two trial states.
\label{fig:correlation}
}
\end{figure}

Our trial wave functions are motivated by the mathematical structure of the theory. To illustrate what physics these wave functions encode we will calculate a few observables in the original basis of electrons and spins. Quantities of special interest are the spin-spin correlation functions
\begin{equation}
\chi_{\alpha \beta}(\mathbf{r}_{ij}) = 
\la \mathbf S_\alpha(\mathbf{r}_i)\cdot \mathbf S_\beta(\mathbf{r}_j) \ra ,
\end{equation}
with $\alpha = c,f$ and $\beta = c,f$. In particular we will consider $\chi_{fc}$, which is a measure of how the $f$-spins and $c$-electrons are entangled with one another, and $\chi_{ff}$.
A straightforward somewhat tedious calculation for the confined state of section~\ref{subsec:trial1}, with three-body bound states and uncorrelated rotated spins, gives
\begin{multline}
\chi_{fc}(\mathbf{r}_{ij}) = 
- \frac{3}{8}\delta_{ij} \Bigl({\frac{1}{N}\sum_{\mathbf{k}}}'  \sin(\alpha_{\mathbf{k}}) \Bigr)
\\
-\frac{3}{8} 
\Bigl(
{\frac{1}{N}\sum_{\mathbf{k}}}'  \sin(\alpha_{\mathbf{k}})
\cos(\mathbf{k} \cdot \mathbf{r}_{ij})
 \Bigr)^2
\\
+ \frac{3}{2}\Bigl({\frac{1}{N}\sum_{\mathbf{k}}}'  \sin(\alpha_{\mathbf{k}}) \Bigr)
\Bigl(
{\frac{1}{N}\sum_{\mathbf{k}}}'  \sin^2(\alpha_{\mathbf{k}}/2)
\sin(\mathbf{k} \cdot \mathbf{r}_{ij})
 \Bigr)^2
 \\
 +
 \frac{3}{4} \sum_{m,n} \mathcal{A}_{in}^2 \mathcal{B}_{jn}  \mathcal{A}_{jm}^3
\Bigl(
{\frac{1}{N}\sum_{\mathbf{k}}}' 
\sin(\mathbf{k} \cdot \mathbf{r}_{mn})
 \Bigr) .
\end{multline}
The first two terms show that the on-site spin-spin correlation function is always $\geq - 9/32$ in this state. The last two terms are only non-zero when the coordinates are on different sublattices. Some insight into the typical behavior can be obtained from the special case of a constant $\alpha_\mathbf{k} = \alpha$, in this case the correlation function becomes
\begin{multline}
\chi_{fc}(\mathbf{r}_{ij}) = 
-  \frac{3}{32}  \sin(\alpha) \bigl[ 2 + \sin(\alpha) \bigr]\delta_{ij}
\\
+ \frac{3}{8} \sin(\alpha) \bigl[ 1+\cos^2(\alpha) \bigr] 
\Bigl(
{\frac{1}{N}\sum_{\mathbf{k}}}' 
\sin(\mathbf{k} \cdot \mathbf{r}_{ji})
 \Bigr)^2 .
\end{multline}
The last term decays as $1/r^2$ in 1D. In the O(3)-symmetric confined state of section~\ref{subsec:trial2} the correlation function is 
\begin{multline}
\chi_{fc}(\mathbf{r}_{ij}) = 
- \Bigl(
{\frac{1}{N}\sum_{\mathbf{k}}}' 
 \sin(\alpha_\mathbf{k})
 \Bigr)\frac{3}{4}\delta_{ij} 
 \\
 + 3 
\Bigl(
{\frac{1}{N}\sum_{\mathbf{k}}}' 
 \sin(\alpha_\mathbf{k})
 \Bigr)
 \Bigl(
{\frac{1}{N}\sum_{\mathbf{k}}}' \cos(\alpha_\mathbf{k})
\sin(\mathbf{k} \cdot \mathbf{r}_{ij})
 \Bigr)^2
 \\
 -\frac{3}{2}  \Bigl(
{\frac{1}{N}\sum_{\mathbf{k}}}' \sin(\alpha_\mathbf{k})
\cos(\mathbf{k} \cdot \mathbf{r}_{ij})
 \Bigr)^2 .
\end{multline}
Specializing to constant $\alpha$ this simplifies to
\begin{multline}
\chi_{fc}(\mathbf{r}_{ij}) = 
-\frac{3}{8} \sin(\alpha) \bigl[ 1 + \sin(\alpha) \bigr] \delta_{ij}
\\
+ \frac{3}{2} \sin(\alpha) \cos^2(\alpha) 
\Bigl(
{\frac{1}{N}\sum_{\mathbf{k}}}' 
\sin(\mathbf{k} \cdot \mathbf{r}_{ji})
 \Bigr)^2 .
\end{multline}
The first term shows that local singlets can be described with this state, taking $\alpha = \pi/2$. The $\chi_{ff}$ correlation function is simpler to be computed, for the confined state it is
\begin{equation}
\chi_{ff}(\mathbf{r}_{ij}) =  \frac{3}{4} \delta_{ij}
- 3 
 \Bigl(
{\frac{1}{N}\sum_{\mathbf{k}}}' \sin^2(\alpha_\mathbf{k}/2)
\sin(\mathbf{k} \cdot \mathbf{r}_{ij})
 \Bigr)^2 ,
\end{equation}
while for the deconfined state it is
\begin{equation}
\chi_{ff}(\mathbf{r}_{ij}) =  \frac{3}{4} \delta_{ij}
- 3 
 \Bigl(
{\frac{1}{N}\sum_{\mathbf{k}}}' \cos(\alpha_\mathbf{k})
\sin(\mathbf{k} \cdot \mathbf{r}_{ij})
 \Bigr)^2 .
\end{equation}
Typical results for the correlation functions are illustrated in Fig.~\ref{fig:correlation}. The correlation functions in the two states are qualitatively similar but the decay lengths and the magnitudes are quantitatively different. Let us also note that the smallness of $\chi_{ff}$ can be \emph{dramatically increased} by including correlations among the rotated spins in the confined state as discussed in Sec.~\ref{sec:variationspincorrelation}. The energy scale associated with the rotated spins is that of the RKKY interactions, i.e., $\sim J^2/t$ up to possibly logarithmic corrections for small $J/t$. 

\begin{figure}
\centering
\includegraphics[scale=.85]{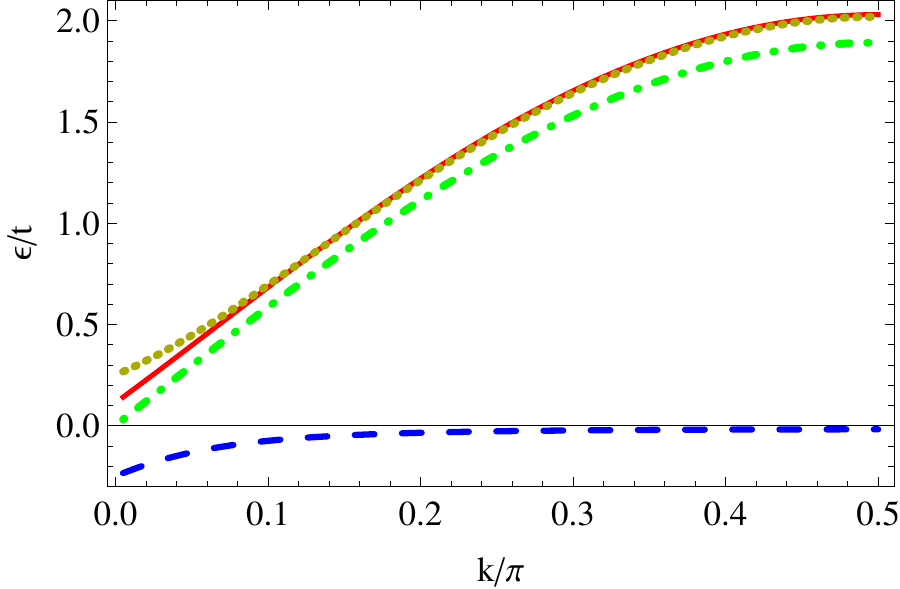}
\caption{
Typical mean-field band structures for the two trial states in 1D, optimized for $J = t$. Confined state:
 $\tilde{\epsilon}_{\gamma,a}(k)$, $a=1,2,3$ (solid red), $\tilde{\epsilon}_{\gamma,0}(k)$ (dash-dotted green); Deconfined state: 
  $\tilde{\epsilon}_{\gamma,a}(k)$, $a=1,2,3$ (dotted yellow),
   $\tilde{\epsilon}_{\mu,a}(k)$, $a=1,2,3$ (dashed blue).
   The gapless  $\tilde{\epsilon}_{\gamma,0}(k)$ band is renormalized by a factor of $\mathcal{A}_{00}^6$ compared to the free band $\epsilon_k$.
\label{fig:MFbands}
}
\end{figure}

We will now briefly discuss the mean-field band structure of the two states, which are illustrated in Fig.~\ref{fig:MFbands}. The deconfined state has two sets of gapped Majorana excitations, both being 3-fold degenerate. The confined state has one set of three-fold degenerate gapped Majorana excitations and one gapless Majorana mode. The latter state is however clearly not the ground state since we can gain energy from the RKKY interaction, which is induced by the transformation, of the rotated spins. Nevertheless we expect that the trial state with uncorrelated rotated spins is a good one in the temperature range $J^2/t \lesssim T \lesssim J/4 $ where thermal fluctuations have made the rotated spin system disordered. In this temperature range we expect that the much of the physics is dominated by the linearly dispersing gapless Majorana mode discussed in Ref.~\onlinecite{Miranda1994}.

In the small $J/t$-limit the ground state $f$-spin correlation function $\chi_{ff}$ is dominated by the RKKY-interaction,\cite{YuWhite1993} which is not described well by the state with uncorrelated rotated spins. This can be improved by allowing for correlations among the rotated spins as discussed in Sec.~\ref{sec:variationspincorrelation}. There are less studies of $\chi_{fc}$ in the literature, but the magnitude of $\chi_{fc}(\mathbf{r}=\mathbf{0})$ in the deconfined state agrees well with the numerical values of Ref.~\onlinecite{YuWhite1993} at $J=t$. Finally we would like to remark that the correlation function  $\chi_{fc}(\mathbf{r})$ of the two states (confined and deconfined) are qualitatively similar. This is consistent with the notion that the two states can be adiabatically connected, i.e., there is no phase transition in at half-filling in 1D in going between small and large $J/t$-limits. \cite{Sigrist_RMP_1997}

\section{Path integral formulation for Majorana fermions}
\label{sec:MajoranaPathIntegral}

Using standard path integral formulas and fermion coherent states,\cite{NO_Book} we can calculate the expectation value for an operator $\hat{\mathcal{O}}$ given the density matrix $\hat{\rho} = e^{-\beta \tilde{H}}/Z_0$ from
\begin{multline}
\text{Tr} \hat{\mathcal{O}} \hat{\rho} = 
\int \Bigl(\prod_\alpha d \xi^*_{\alpha,M} d \xi^{\,}_{\alpha,0}  d \xi^*_{\alpha,0} d \xi^{\,}_{\alpha,1} \Bigr) \la -\xi_0 |\hat{\mathcal{O}} | \xi_0 \ra
\\ \times
e^{-\sum_\alpha ( \xi^*_{\alpha,M} \xi^{\,}_{\alpha,0}+ \xi^*_{\alpha,0} \zeta^{\,}_{\alpha,1})} 
\la  \xi_M | \hat{\rho} | \xi_1 \ra   . 
\end{multline}
$Z_0 =e^{S_0}$ is a constant that is needed to insure that $\text{Tr} \hat{\rho} = 1$.
The formalism can also easily be adapted to the calculation of the partition function, in which case we substitute $\hat{\rho} \rightarrow e^{-\beta \hat{H}} $ and set $\hat{\mathcal{O}} = 1$. Now we write $\hat{\rho} = \hat{\rho}^{1/M} \hat{\rho}^{1/M} \cdots \hat{\rho}^{1/M}$ and insert $M-1$ identity operators with the slightly non-standard convention
\begin{equation}
1 =  \int \Bigl( \prod_{\alpha} d \xi^{*}_{\alpha,l} d \xi^{\,}_{\alpha,l+1} \Bigr)
e^{-\sum_\alpha \xi^{*}_{\alpha,l} \xi^{\,}_{\alpha,l+1}} 
| \xi_{l+1} \ra \la \xi_{l}| .
\end{equation}
The standard 
procedure then gives ($\Delta \tau = \beta / M$)
\begin{multline}
\la  \xi_M | \hat{\rho} | \xi_1 \ra \propto 
e^{-\sum_\alpha \xi^{*}_{\alpha,M} \xi^{\,}_{\alpha,1}} 
e^{\sum_{\alpha,j}( 1-e^{-i \omega_j})\xi^{*}_{\alpha,\omega_j} \xi^{\,}_{\alpha,\omega_j} }
\\ \times
e^{-\sum_l \Delta \tau \tilde{H}(\xi^{*}_l, \xi^{\,}_l) }
=
e^{-\sum_\alpha \xi^{*}_{\alpha,M} \xi^{\,}_{\alpha,1}} e^{-S(\xi^{*},\xi)},
\label{eq:rhomatrixelement1}
\end{multline}
up to corrections of order $1/M$. Here we have also used the discrete Fourier convention (with $l = 1 , \ldots , M$) 
\begin{equation}
\xi^{\,}_{\alpha,l} = \frac{1}{\sqrt{M}}\sum_{j} e^{-i \omega_j l} \xi^{\,}_{\alpha,\omega_j}
, \quad
\xi^*_{\alpha,l} = \frac{1}{\sqrt{M}}\sum_{j} e^{i \omega_j l} \xi^*_{\alpha,\omega_j}.
\end{equation}
The $j$-sum is over the $M$ distinct solutions to $e^{i \omega_j M} = -1$, corresponding to anti-periodic boundary conditions. We will take $M$ to be even so that all positive frequencies can be paired up with a corresponding negative one $\omega_{-j} = -\omega_j$. This is a unitary transformation and hence the corresponding Jacobian is just a phase. The proportionality sign in \eqref{eq:rhomatrixelement1} is needed since we have left out the integration over the internal variables.  After rearrangement the full integration measure becomes 
\begin{multline}
\int d[\xi] = 
 \int 
 \Bigl( \prod_{\alpha}  d \xi^{*}_{\alpha,0} d \xi^{\,}_{\alpha,0}   \Bigr)
  \Bigl( \prod_{\alpha,l} d \xi^{\,}_{\alpha,l}  d \xi^{*}_{\alpha,l}  \Bigr)
\\  =
 \int 
 \Bigl( \prod_{\alpha} d \xi^{*}_{\alpha,0}  d \xi^{\,}_{\alpha,0} \Bigr)
 \Bigl( \prod_{\alpha,j} d \xi^{\,}_{\alpha,\omega_j} d \xi^{*}_{\alpha,\omega_j}\Bigr),
\end{multline}
and the path integral representation for the trace is
\begin{multline}
\label{eq:integranden1}
\text{Tr} \hat{\mathcal{O}} \hat{\rho} = 
\int d[\xi]
e^{-\sum_\alpha ( \xi^{*}_{\alpha,M} +  \xi^{*}_{\alpha,0}) 
(\xi^{\,}_{\alpha,1}+ \xi^{\,}_{\alpha,0} )} 
\\ \times \mathcal{O}(-\xi^*_0,\xi^{\,}_0) e^{-S(\xi^* , \xi)} e^{-S_0}.
\end{multline}
A matrix $\tilde{H}$ that allows for a frequency dependent self-energy is obtained via the substitution
\begin{equation}
\sum_l \Delta \tau \tilde{H}(\xi^{*}_l, \xi^{\,}_l) \longrightarrow \frac{\beta}{M}
\sum_\alpha \sum_j \epsilon_\alpha(\omega_j) \xi^{*}_{\alpha,\omega_j} \xi^{\,}_{\alpha,\omega_j},
\end{equation}
whereas for a diagonal quadratic Hamiltonian we will have $ \epsilon_\alpha(\omega_j) =  \epsilon_\alpha$, independent of $\omega_j$. Introducing the propagator via the relation $G^{-1}_\alpha (\omega_j) = -\bigl[ 1-e^{-i \omega_j}  -\beta \epsilon_\alpha(\omega_j)/M\bigr]$ we may then write the quadratic part of the action as
\begin{equation}
\label{eq:FrequencyDependentAction}
S(\xi^* , \xi) = \sum_{\alpha,j} \xi_{\alpha,\omega_j}^* G^{-1}_\alpha (\omega_j) \xi_{\alpha,\omega_j}^{\,} .
\end{equation}
This implies that the action is diagonal in the $\alpha$-index, i.e., we can write $\hat{\rho} = \prod_\alpha \hat{\rho}_\alpha$. All of these are well-known results, let us now adapt this formalism to Majorana operators.

\subsection{Majorana path integral}

Let us now consider a Hamiltonian that is defined in terms of a set of $N_\chi$ Majorana fermion operators $\hat{\chi}_\alpha$. To use standard fermion coherent states the Majoranas have to be paired up to form conventional Dirac fermions. There are two straightforward ways to do this, called ``fermion doubling'' and ``fermion halving'' in Ref.~\onlinecite{deBoer1996}. In fermion halving the $\hat{\chi}_\alpha$ themselves are paired up, this procedure is outlined in Ref.~\onlinecite{ShankarVishwanath2011}. We use fermion doubling instead, which is more symmetric and relies on an extension of the Hilbert space. For each $\hat{\chi}_\alpha$ we introduce a conventional fermion annihilation operator $c_\alpha$ and represent $\hat{\chi}_\alpha$ by $\frac{1}{\sqrt{2}}(c_\alpha^{\,} + c_\alpha^{\dagger})$. The other Majorana operators $ \frac{i}{\sqrt{2}}(c_\alpha^{\,} - c_\alpha^{\dagger})$ do not enter the Hamiltonian and can be paired up to form an independent Hilbert space that will generate an overall degeneracy factor of $2^{N_\chi /2}$ in the trace. To avoid fractional degeneracy we will take $N_\chi$ to be even, this is also always the case in physical situations where Majorana fermions unconditionally appear in pairs. This construction means that we have artificially doubled the size of the Hilbert space; we will eliminate these extra degrees of freedom from the path integral later.
We can now directly use the path integral expressions above and perform a basis change for the Grassmann numbers to
\begin{equation}
\begin{split}
\zeta_{\alpha,l} = \frac{\xi^{\,}_{\alpha,l} + \xi^{*}_{\alpha,l}}{\sqrt{2}}
, \qquad
\nu_{\alpha,l} = \frac{ \xi^{\,}_{\alpha,l} - \xi^{*}_{\alpha,l} }{\sqrt{2}}
, \\ 
d  \xi^{\,}_{\alpha,l} d  \xi^{*}_{\alpha,l}  =   d  \nu_{\alpha,l} d  \zeta_{\alpha,l},
\end{split}
\end{equation}
for  $l=1, \ldots , M$, and
\begin{equation}
\begin{split}
\zeta_{\alpha,0} =  \frac{\xi^{\,}_{\alpha,0} - \xi^{*}_{\alpha,0}}{\sqrt{2}}
,  \qquad
\nu_{\alpha,0} =  \frac{ \xi^{\,}_{\alpha,0} + \xi^{*}_{\alpha,0} }{\sqrt{2}}
, \\ 
d \xi^{*}_{\alpha,0} d \xi^{\,}_{\alpha,0}   =  d  \nu_{\alpha,0} d \zeta_{\alpha,0}.
\end{split}
\end{equation}
The integrand of \eqref{eq:integranden1} then goes into
\begin{equation}
\label{eq:integranden2}
e^{-\sum_\alpha 
( \xi^{*}_{\alpha,M} +  \xi^{*}_{\alpha,0}) 
(\xi^{\,}_{\alpha,1}+ \xi^{\,}_{\alpha,0} )}
 \mathcal{O}(\zeta^{\,}_0) e^{-S(\zeta,\nu)} e^{-S_0} .
\end{equation}
Expanding the new Grassmann variables as $\zeta_{\alpha,l} = \frac{1}{\sqrt{M}}\sum_j e^{-i l \omega_j} \zeta_{\alpha,\omega_j}$ (and the same for $\nu_{\alpha,l}$) the action becomes
\begin{eqnarray}
S(\zeta,\nu) &=& \sum_l \Delta \tau \tilde{H}(\zeta_l)
\nonumber \\
&+& \sum_\alpha {\sum_{j}}' 
\bigl[ 1 - \cos(\omega_j) \bigr] 
(\nu_{\alpha,-\omega_j} \zeta_{\alpha,\omega_j} - \zeta_{\alpha,-\omega_j} \nu_{\alpha,\omega_j}  )
\nonumber \\
&+ & \sum_\alpha {\sum_{j}}' 
i \sin(\omega_j) 
(\nu_{\alpha,-\omega_j} \nu_{\alpha,\omega_j} - \zeta_{\alpha,-\omega_j} \zeta_{\alpha,\omega_j} ).
\end{eqnarray}
Here the prime on the sum means that only positive frequencies are to be included.
Expanding out a generic $\mathcal{O}(\zeta_0)$ we have (repeated indexes are summed over)
\begin{multline}
\mathcal{O}(\zeta_0) = \zeta_{\alpha,0} H^{(2)}_{\alpha \alpha'} \zeta_{\alpha',0}
+  H^{(4)}_{\alpha\beta \gamma \delta}  \zeta_{\alpha,0} \zeta_{\beta,0} \zeta_{\gamma,0} \zeta_{\delta,0}
\\ 
+
H^{(6)}_{\alpha\beta\gamma\alpha' \beta' \gamma'}  
\zeta_{\alpha,0} \zeta_{\beta,0}  \zeta_{\gamma,0}
\zeta_{\alpha',0}  \zeta_{\beta',0} \zeta_{\gamma',0}
+ \ldots
\label{eq:MajoranaHamiltonianExpansion}
\end{multline}
The integral over variables $\xi^{\,}_{\alpha,0}$ and  $\xi^{*}_{\alpha,0}$ for $\alpha$'s that are not present in the expansion of $\mathcal{O}(\zeta_0)$ gives one together with the prefactor, i.e.,
\begin{equation}
\int  d  \xi^{*}_{\alpha,0} d \xi^{\,}_{\alpha,0} 
e^{-
 ( \xi^{*}_{\alpha,M} +  \xi^{*}_{\alpha,0}) 
(\xi^{\,}_{\alpha,1}+ \xi^{\,}_{\alpha,0} )
} =1.
\end{equation}
For $\alpha$'s that are present in $\mathcal{O}(\zeta_0)$ the prefactor can be replaced by
\begin{equation}
e^{ -( \xi^{*}_{\alpha,M} +  \xi^{*}_{\alpha,0}) 
(\xi^{\,}_{\alpha,1}+ \xi^{\,}_{\alpha,0} )
}
\rightarrow
e^{-\nu_{\alpha,0} (\xi^{\,}_{\alpha,1} - \xi^{*}_{\alpha,M} )/\sqrt{2}} .
\end{equation}
Expanding this and performing the integral over $\zeta_{0,\alpha}$ and $\nu_{\alpha,0}$ implies that we can substitute
\begin{multline}
\zeta_{\alpha,0} \longrightarrow 
-
\frac{\xi^{\,}_{\alpha,1} - \xi^{*}_{\alpha,M} }{\sqrt{2}}
= -
\frac{1}{\sqrt{M}} \sum_j e^{-i \omega_j /2}
\\ \times
\left[
\cos(\omega_j / 2)\zeta_{\alpha,\omega_j} - i \sin(\omega_j / 2) \nu_{\alpha,\omega_j} \right],
 \end{multline}
in $\mathcal{O}(\zeta_0)$. This implies that the integrand goes into
\begin{eqnarray}
\mathcal{O} \Bigl( \frac{\xi^{*}_{\alpha,M}  -\xi^{\,}_{\alpha,1} }{\sqrt{2}} \Bigr) e^{-S(\zeta,\nu)}e^{-S_0}.
 \end{eqnarray}
The corresponding integral (average) is easily evaluated by taking derivatives of the generating function with respect to the Grassmann source fields $\eta_\alpha$:
 \begin{equation}
Z(\eta) =2^{-N_\chi /2} e^{-S_0} \int d[\xi]
e^{-S(\zeta,\nu)} e^{\sum_\alpha \eta_{\alpha} \frac{ \xi^{*}_{\alpha,M}  - \xi^{\,}_{\alpha,1}}{\sqrt{2}} },
 \end{equation}
where the measure can be written as
\begin{equation}
d[\xi] = \prod_\alpha {\prod_j}' 
d \nu_{\alpha,-\omega_j} d \nu_{\alpha,\omega_j}
d \zeta_{\alpha,-\omega_j} d \zeta_{\alpha,\omega_j}
= d[\nu] d[\zeta] .
\end{equation}
Because the exponent is now a quadratic polynomial in the $\nu$'s, these variables can be integrated out exactly. Therefore the generating function is given in terms of an integral over Grassmann variables obtained by replacing, locally in the action, the Majorana operators in the Hamiltonian with Grassmann numbers. The imaginary time evolution of these numbers deviate from those of conventional fermions. In fact, after some straightforward algebra, the result is (dropping the overall normalization factor)
\begin{equation}
\label{eq:GrassmannMajoranaAction1}
Z(\eta) \propto
\int d[\zeta] e^{-S(\zeta,\eta)} ,
\end{equation}
\begin{multline*}
S(\zeta,\eta) = \Delta \tau \sum_l \tilde{H}(\zeta_l)
- i  \,{\sum_{j,\alpha}} \tan\left(\frac{\omega_j}{2}\right)
 \zeta_{\alpha,-\omega_j} \zeta_{\alpha,\omega_j} 
\\
+
\sum_\alpha  \frac{\eta_\alpha}{\sqrt{M}}
{\sum_j} 
\frac{2  \zeta_{\alpha,\omega_j} }{1+e^{i \omega_j}}.
 \end{multline*}
This result can be simplified further by a linear transformation to another set of Grassmann numbers $\chi$, defined via 
\begin{equation}
\chi_{\alpha,\omega_j} \equiv 
\frac{e^{-i \omega_j /2}  \zeta_{\alpha,\omega_j} }{\cos(\omega_j /2)},
\end{equation}
or equivalently, in discrete imaginary time
\begin{equation}
\zeta_{\alpha,l} = \frac{1}{2} \left[ \chi_{\alpha,l} + (1-\delta_{l,M}) \chi_{\alpha,l+1} - \delta_{l,M} \chi_{\alpha,1} \right] .
\label{eq:midpointrule}
\end{equation}
In terms of these variables the generating function is
\begin{equation}
Z(\eta) = 2^{N_\chi (M-1)/2} \int d[\chi] e^{-S(\chi,\eta)},
\label{eq:GrassmannMajoranaAction}
\end{equation}
\begin{multline*}
S(\chi,\eta) = \Delta \tau \sum_l \tilde{H}(\zeta_l)
-
 i  \,{\sum_j} \frac{\sin(\omega_j)}{2} 
 \chi_{\alpha,-\omega_j} \chi_{\alpha,\omega_j} 
\\ +\sum_\alpha
\frac{ \eta_\alpha}{\sqrt{M}} \sum_j  \chi_{\alpha,\omega_j}.
 \end{multline*}
In this expression it is clear that the source term is localized exactly at time zero. The main result of this section is that it is possible to reduce the calculation of the partition function (or a density matrix) for a Hamiltonian that is expressed in terms of Majorana fermions, to a standard Grassmann integral obtained by substituting the Majorana operators $\hat{\chi}_{\alpha}$ with Grassmann numbers $\chi_{\alpha,l}$ for each time slice $l$. The properly regularized  generating function is then given by \eqref{eq:GrassmannMajoranaAction1} or \eqref{eq:GrassmannMajoranaAction}. The corresponding continuum short-hand expression is (cf. Refs.~\onlinecite{Tsvelik_book,ShankarVishwanath2011})
\begin{multline}
S(\chi,\eta) = \int_0^\beta d \tau \Bigl(  
\frac{1}{2}\sum_\alpha  \chi_\alpha(\tau) \partial_\tau \chi_\alpha (\tau)+ H[\chi(\tau)]
\\  + \sum_\alpha \eta_\alpha(\tau)  \chi_\alpha(\tau) \Bigr),
\end{multline}
where we have also included source fields at intermediate times. The regularization in \eqref{eq:GrassmannMajoranaAction} is a kind of ``midpoint rule'', \cite{deBoer1996} which is clear from \eqref{eq:midpointrule}. In  \eqref{eq:GrassmannMajoranaAction} we have also included the correct prefactor $2^{N_\chi (M-1)/2}$ that is necessary to get the actual value of the partition function. This can be derived by considering the partition function for $\tilde{H} = \eta_\alpha =0$, or alternatively, by carefully keeping track of the prefactor generated by the elimination of the $\nu$'s and the change in the measure when going from $\int d[\zeta]$ to $\int d[\chi]$. We finally note that the Grassmann integral for a quadratic action that is not time-translational invariant can in general be expressed in terms of a Pfaffian, see for example Refs.~\onlinecite{Stone_Book_QFT,Robledo2009}.

\subsection{Consistency check}

As a consistency check we apply the formalism of the last section to the simplest Majorana Hamiltonian $\hat{H} = - i \epsilon \hat{\chi}_1 \hat{\chi}_2$ and compute $\la 2 i \hat{\chi}_1 \hat{\chi}_2 \ra$. In terms of the standard fermion annihilation operator $\hat{c} = (\hat{\chi}_1 -i \hat{\chi}_2)/\sqrt{2}$ the Hamiltonian is $\hat{H} = \epsilon (\hat{c}^\dagger \hat{c}-1/2)$ and $2 i \hat{\chi}_1 \hat{\chi}_2 = 1 - 2 \hat{c}^\dagger \hat{c}$. A simple finite temperature calculation then gives $\la 1 - 2 \hat{c}^\dagger \hat{c} \ra = \tanh(\epsilon \beta /2)$. Let us now reproduce this result using the Majorana path integral. Taking derivatives of the generating function and setting the source fields to zero we obtain
\begin{eqnarray}
\la 2 i \hat{\chi}_1 \hat{\chi}_2 \ra = 
\frac{2i}{M} \sum_j \frac{ \int d[\chi ]\chi_{1,-\omega_j} \chi_{2,\omega_j} e^{-S(\chi,0)}}{Z(0)},
 \end{eqnarray}
where in this case
\begin{multline}
\label{eq:simplestaction}
S(\chi,0) = \frac{-i \epsilon \beta}{M} \sum_j \chi_{1,-\omega_j} \chi_{2,\omega_j} 
\cos^2(\omega_j /2)
\\
-i \sum_\alpha {\sum_j}' \sin(\omega_j)
 \chi_{\alpha,-\omega_j} \chi_{\alpha,\omega_j} ,
\end{multline}
so that
\begin{eqnarray}
\la 2 i \hat{\chi}_1 \hat{\chi}_2 \ra = 
\frac{2 a}{M} \sum_j
\frac{1}{4\sin^2(\omega_j /2) + \cos^2(\omega_j /2) a^2},
\label{eq:summation1}
\end{eqnarray}
where $a = \beta \epsilon/M$. To perform frequency summations we use the following standard trick: if $f(z)$ is a function that is regular at the points $z=e^{-i \omega_j}$ and sufficiently well-behaved at infinity we have the relation
\begin{equation}
 \frac{1}{M} \sum_{j} e^{-i \omega_j} f(e^{-i \omega_j}) =
 \oint_C \frac{dz}{2 \pi i} \frac{f(z)}{z^M + 1} ,
 \label{eq:summationformula}
\end{equation}
which is often easily evaluated using residues. The contour $C$ should enclose all of the singularities of $f(z)$ but not those of $(z^M+1)^{-1}$. Using this, the sum in \eqref{eq:summation1} is easily evaluated:
\begin{equation}
\la 2 i \hat{\chi}_1 \hat{\chi}_2 \ra = 
-2a \oint_C \frac{dz}{2 \pi i} \frac{1}{(z^M+1)}
\frac{1}{(z-1)^2 - (z+1)^2 (a/2)^2} .
\end{equation}
The residues are located at $z_1 = (1+a/2)/(1-a/2)$ and $z_2 = 1/z_1$, adding up their contribution we arrive at
\begin{multline}
\la 2 i \hat{\chi}_1 \hat{\chi}_2 \ra =-
\left( \frac{1}{(z_1^M+1)} - \frac{1}{(z_2^M+1)} \right)
\\
\rightarrow
\left( \frac{1}{e^{-\beta \epsilon}+1} - \frac{1}{e^{\beta \epsilon}+1}  \right)
= \tanh(\beta \epsilon/2),
\end{multline}
which is the correct result.

\section{Application of the Majorana path integral to the Kondo lattice}
\label{sec:trialdensitymatrix}

In this section we generalize the first (favored for small $J/t$) variational calculation of section~\ref{sec:operatorvariation} to finite temperatures in terms of a path integral representation of a trial density matrix. Of particular interest is how to describe the formation of the composite object $\tilde{\gamma}_0 = 2 i \tilde{\mu}_1 \tilde{\mu}_2 \tilde{\mu}_3$ in the path integral language. Let us first recall the finite temperature variational principle,\cite{Feynman_SM,Chaikin_Lubensky_Book} which states that the trial free energy $F_{\hat{\rho}}$ is an upper bound to the true free energy $F$ for every properly normalized density matrix $\hat{\rho}$, with
\begin{equation}
F_{\hat{\rho}} = \text{Tr} \hat{H} \hat{\rho}  + T \, \text{Tr}  \hat{\rho} \log \hat{\rho}  \geq F.
\label{eq:trial_free_energy}
\end{equation}
In our case we will parametrize $\hat{\rho}$ through its matrix elements between coherent states in the rotated basis
\begin{eqnarray}
\la  \tilde{\xi}_M | \hat{\rho} | \tilde{\xi}_1 \ra 
&\propto& e^{-\sum_{a=0}^3 S_{\tilde{\gamma}_a}}
\nonumber  \\
&\times&
e^{-\sum_{a=1}^3 \sum_{\mathbf{k}}' \left[ \tilde{\gamma}^{*}_{a,M} (\mathbf{k}) \tilde{\gamma}^{\,}_{a,1} (\mathbf{k}) 
+ \tilde{\mu}^{*}_{a,M} (\mathbf{k}) \tilde{\mu}^{\,}_{a,1} (\mathbf{k}) \right] } ,
\nonumber  \\
S_{\tilde{\gamma}_a} &=&  \sum_j {\sum_{\mathbf{k}}}' 
G_{\tilde{\gamma}_a}^{-1}(\omega_j,\mathbf{k}) {\tilde{\gamma}}^*_a (\omega_j,\mathbf{k}) 
\tilde{\gamma}^{\,}_a (\omega_j,\mathbf{k}) ,
\nonumber \\
G_{\tilde{\gamma}_a}^{-1}(\omega_j,\mathbf{k}) &=&
-(1-e^{-i \omega_j}  -\beta E_{\tilde{\gamma}_a}(\mathbf{k})/M) .
\label{eq:trialdensitymatrixparametrization}
\end{eqnarray}
The variational parameters are $\alpha_{\mathbf{k}}$, which define the operators $\tilde{\gamma}^{\,}_{a} (\mathbf{k})$ and $\tilde{\mu}^{\,}_{a} (\mathbf{k})$ via \eqref{eq:Majoranarotation1}, and the trial energies in the propagators $E_{\tilde{\gamma}_a}(\mathbf{k})$.

\subsection{Energy expectation value}

The energy expectation values of $\tilde{\gamma}$'s are easily calculated using the formula
\begin{equation}
\text{Tr}  \hat{n}_\alpha \hat{\rho} = 
1 - \left( \frac{1}{M} \sum_{j}  e^{-i \omega_j} G_{\alpha} (\omega_j) \right)
= \frac{1}{1+e^{\beta E_\alpha}}  \equiv \bar{n}_\alpha,
\end{equation}
where $\alpha$ is a short-hand for $(\tilde{\gamma}_{a}, \mathbf{k})$. This formula is obtained from the path integral using the free fermion propagator of \eqref{eq:trialdensitymatrixparametrization}, with the aid of \eqref{eq:summationformula} in the limit $M \rightarrow \infty$. Using this we find
\begin{equation}
\text{Tr}  \hat{H}^{(2)} \hat{\rho} = \frac{3}{4} E_{0} + \frac{3}{2} {\sum_{\mathbf{k}}}' \tilde{\epsilon}_\mu (\mathbf{k}) 
+
\sum_{a=1}^3 {\sum_{\mathbf{k}}}' \tilde{\epsilon}_\gamma(\mathbf{k}) \bar{n}_{\tilde{\gamma}_a} (\mathbf{k}) 
 ,
\end{equation}
and
\begin{multline}
\text{Tr}  \hat{H}^{(4)} \hat{\rho} = - \frac{J}{2 N}
\Bigl( {\sum_{\mathbf{k}}}' \sin(\alpha_{\mathbf{k}}) \Bigl[ \bar{n}_{\tilde{\gamma}_1} (\mathbf{k}) -\frac{1}{2} \Bigr] \Bigr)
\\ \qquad \qquad \qquad \times
\Bigl( {\sum_{\mathbf{k}}}' \sin(\alpha_{\mathbf{k}}) \Bigl[ \bar{n}_{\tilde{\gamma}_2} (\mathbf{k}) -\frac{1}{2} \Bigr] \Bigr)
\\
+  {\rm \; cyclic \; permutations}\; 1\rightarrow 2 \rightarrow 3 \rightarrow 1 .
\end{multline}
To arrive at these results we have also performed the average involving $\tilde{\mu}$'s. We will now describe how this can be done.

\subsection{Bound state formation and the six-fermion term}

The average of terms involving $\tilde{\mu}$'s can be formulated as integrals of the form
\begin{multline}
\int \left( \prod_{i,l} 
d \tilde{\mu}_{1,l}^{\,}( \mathbf{r}_i)  d\tilde{\mu}_{2,l}^{\,}( \mathbf{r}_i)  d\tilde{\mu}_{3,l}^{\,}( \mathbf{r}_i)
\right)
e^{-S_{\tilde{\gamma}_0}} H(\tilde{\mu}_0)
 \\ = \int d[ \tilde{\mu}] e^{-S_{\tilde{\gamma}_0}} H(\tilde{\mu}_0).
\end{multline}
Here $H(\tilde{\mu}_0)$ is a short-hand for a term in the Hamiltonian where the operators $\tilde{\mu}_\alpha$ have been substituted with the corresponding Grassmann number of the zeroth time slice $\tilde{\mu}_{\alpha,0}$, cf. \eqref{eq:MajoranaHamiltonianExpansion}. The Grassmann numbers in \eqref{eq:trialdensitymatrixparametrization} can be expressed in terms of the real space and imaginary time as
\begin{eqnarray}
\tilde{\gamma}^{\,}_{0,\omega_j} (\mathbf{k}) &=&
 \frac{1}{\sqrt{N M}} \sum_{\mathbf{r}_i , j}
e^{i (\omega_j l - \mathbf{k} \cdot \mathbf{r}_i )} \tilde{\gamma}_{0,l}^{\,}( \mathbf{r}_i) ,
\nonumber \\
\tilde{\gamma}^{*}_{0,\omega_j} (\mathbf{k}) &=&
 \frac{1}{\sqrt{N M}} \sum_{\mathbf{r}_i , j}
e^{-i (\omega_j l - \mathbf{k} \cdot \mathbf{r}_i )} \tilde{\gamma}_{0,l}^{\,}( \mathbf{r}_i) ,
\\
\tilde{\gamma}_{0,l}^{\,}( \mathbf{r}_i) &=& 
2 i \tilde{\mu}_{1,l}^{\,}( \mathbf{r}_i)  \tilde{\mu}_{2,l}^{\,}( \mathbf{r}_i)  \tilde{\mu}_{3,l}^{\,}( \mathbf{r}_i).
\nonumber
\end{eqnarray}
This form of the action implies that we are gluing together the rotated Majorana fermions locally in space and imaginary time. To perform the integral we first make use the Grassmann Gaussian integral identity (i.e., a fermionic Hubbard-Stratonovich transformation)
\begin{widetext}
\begin{equation}
e^{-S_{\tilde{\gamma}_0}} = \frac{1}{\det[G_{\tilde{\gamma}_0}]} \int d[\xi]
e^{ \sum_j \sum_{\mathbf{k}}' 
\bigl[ \xi_{\omega_j}^* (\mathbf{k}) G_{\tilde{\gamma}_0}(\omega_j,\mathbf{k}) \xi^{\,}_{\omega_j} (\mathbf{k}) 
+
\xi^{*}_{\omega_j}(\mathbf{k}) \tilde{\gamma}^{\,}_{0,\omega_j}(\mathbf{k})
+
 {\tilde{\gamma}}^*_{0,\omega_j} (\mathbf{k}) 
  \xi^{\,}_{\omega_j} (\mathbf{k})  \bigr]} .
\end{equation}
\end{widetext}
We then write the coupling between the auxiliary Grassmann variables $\xi$ and the $\tilde{\gamma}_0$ as a sum of terms that are local in space and imaginary time
\begin{multline}
\sum_j {\sum_{\mathbf{k}}}' 
\bigl[
\xi_{\omega_j}^* (\mathbf{k}) \tilde{\gamma}^{\,}_{0,\omega_j} (\mathbf{k})
+
 {\tilde{\gamma}}^*_{0,\omega_j} (\mathbf{k})  \xi_{\omega_j} (\mathbf{k}) \bigr]
\\ =
 \sum_{i,l}
 \bigl[ \xi^*_l (\mathbf{r}_i) - \xi^{\,}_l (\mathbf{r}_i) \bigr]
 {\tilde{\gamma}}^{\,}_{0,l} (\mathbf{r}_i) .
\end{multline}
Before performing the integrals over $\tilde{\mu}$ we use another Hubbard-Stratonovich identity to rewrite
\begin{equation}
e^{\xi \tilde{\gamma}_0} = e^{2i \tilde{\mu}_1 \tilde{\mu}_2 \xi \tilde{\mu}_3}
=
\frac{1}{2}\sum_{s=\pm} e^{s (2i \tilde{\mu}_1 \tilde{\mu}_2 + \xi \tilde{\mu}_3)}.
\end{equation}
Here $s$ is an auxiliary field that partly has an interpretation in terms of a fluctuating local magnetic field along the third axis, since it couples to $i \tilde{\mu}_1 \tilde{\mu}_2$. Now we use this identity at each point in space and imaginary time, so that we will have an $s_l(\mathbf{r}_i)$ at each point. It is also convenient to perform a local $Z_2$ gauge transformation to absorb the sign into $\tilde{\mu}_{3,l}(\mathbf{r}_i)$ which will lead to an extra sign in the measure given by the product of all signs $\prod_{l,i} s_l(\mathbf{r}_i)$. We are now in the position to perform the integral over $\tilde{\mu}_1$ and $\tilde{\mu}_2$. Let us expand $H(\tilde{\mu}_0)$ in position and time. The most important six-fermion terms coming from \eqref{eq:H6} are
\begin{widetext}
\begin{multline}
H^{(6)}(\tilde{\mu}_0) \sim it (2i)^2 \sum_{i,d} \sum_{j,k,l} \sum_{j',k',l'} 
\mathcal{A}_{ij} \mathcal{A}_{ik} \mathcal{A}_{il}
\mathcal{A}_{ij'} \mathcal{A}_{ik'} \mathcal{A}_{il'}
 s_{0}(\mathbf{r}_l + \hat{\mathbf{x}}_d) s_{0}(\mathbf{r}_{l'})  
\\ \times
 \tilde{\mu}_{1,0}(\mathbf{r}_j + \hat{\mathbf{x}}_d)
 \tilde{\mu}_{2,0}(\mathbf{r}_k + \hat{\mathbf{x}}_d)
 \tilde{\mu}_{3,0}(\mathbf{r}_l + \hat{\mathbf{x}}_d)
  \tilde{\mu}_{1,0}(\mathbf{r}_{j'}) \tilde{\mu}_{2,0}(\mathbf{r}_{k'}) \tilde{\mu}_{3,0} (\mathbf{r}_{l'}) .
\end{multline}
\end{widetext}
We note that because of the structure of the transformation with $\mathcal{A}$ and $\mathcal{B}$, the coordinates of the first three Grassmann numbers can not coincide with those of the last three, cf. the discussion below \eqref{eq:transformationmatrices2}. Performing the integral over all the $\tilde{\mu}_1$'s and $\tilde{\mu}_2$'s and summing over all auxiliary fields $s_l(\mathbf{r}_i)$, we see that the result is nonzero only when all of the coordinates coincide, i.e., when $j = k = l$ and $j' = k' = l'$. For each point in space and time we then get a factor of $2$, which can be interpreted as coming from the spin degeneracy. After these manipulations the coupling between $\xi$ and $\tilde{\mu}_3$ in the exponent is simply
\begin{multline}
\sum_{i,l}
\bigl[ \xi^*_l (\mathbf{r}_i) - \xi^{\,}_l (\mathbf{r}_i) \bigr]
{\tilde{\mu}}^{\,}_{3,l} (\mathbf{r}_i) 
 \\ =
\sum_j {\sum_{\mathbf{k}}}' 
\bigl[
\xi_{\omega_j}^* (\mathbf{k}) \tilde{\mu}^{\,}_{3,\omega_j} (\mathbf{k})
+
 {\tilde{\mu}}^*_{3,\omega_j} (\mathbf{k})  \xi_{\omega_j} (\mathbf{k}) \bigr].
\end{multline}
Integrating out $\xi$ and $\xi^*$ we are then back to a simple action for $\tilde{\mu}_3$
\begin{equation}
S_{\tilde{\gamma}_0} \longrightarrow \sum_j {\sum_{\mathbf{k}}}' 
G_{\tilde{\gamma}_0}^{-1}(\omega_j,\mathbf{k}) {\tilde{\mu}}^*_3 (\omega_j,\mathbf{k}) 
\tilde{\mu}^{\,}_3 (\omega_j,\mathbf{k}), 
\label{eq:Sgamma0effective3}
\end{equation}
and the 6-fermion term in question in the Hamiltonian should be replaced by
\begin{equation}
H^{(6)}(\tilde{\mu}_0) \longrightarrow i t \sum_{i,d} \sum_{j,l} 
\mathcal{A}^3_{ij}
\mathcal{A}^3_{il}
\tilde{\mu}_{3,0}(\mathbf{r}_j + \hat{\mathbf{r}}_d)
\tilde{\mu}_{3,0} (\mathbf{r}_{l}) .
\label{eq:H6tildelocal}
\end{equation}
Performing the remaining integral over $\int d[\tilde{\mu}_3]$ we generate the first term of $\la \hat{H}^{(6)} \ra$ in \eqref{eq:H6average1}, the second term is generated from the rotation with six $\mathcal{B}$'s obtained from the original $\hat{H}^{(6)}$ of \eqref{eq:H6}. The average of all other terms in the rotated Hamiltonian involving $\tilde{\mu}$'s will vanish. Physically this is due to the entirely local spin correlations of \eqref{eq:localspincorrelations} that is implicit in the trial density matrix.

Finally we would like to point out that the result of \eqref{eq:Sgamma0effective3}-\eqref{eq:H6tildelocal} can also be derived by considering the fact that the Grassmann integral picks up the coefficient of the term in the integrand where all Grassmann numbers are present. The expansion of $S_{\tilde{\gamma}_0}$ always generates terms where the coordinates of the $\tilde{\mu}$'s go together, since it is an expansion of terms like $\tilde{\gamma}_{0,l}^{\,}( \mathbf{r}_i) = 
2 i \tilde{\mu}_{1,l}^{\,}( \mathbf{r}_i)  \tilde{\mu}_{2,l}^{\,}( \mathbf{r}_i)  \tilde{\mu}_{3,l}^{\,}( \mathbf{r}_i)$. This immediately implies that the coordinates of $H^{(6)}(\tilde{\mu}_0)$ has to be paired up accordingly for the result to be non-zero. The virtue of the formalism of this section is that it can be used also in other contexts where the Hamiltonian is not so simple. The presented Hubbard-Stratonovich decoupling schemes will be useful in e.g. Quantum Monte Carlo calculations.

\subsection{Entropy calculation}

We would now like to calculate the entropy $S_{\hat{\rho}} = - \text{Tr} \hat{\rho} \ln \hat{\rho}$ for our trial density matrix, which is parametrized in terms of \eqref{eq:trialdensitymatrixparametrization}. A typical term in the action is, with the short-hand notation $\alpha = (\tilde{\gamma}_{a}, \mathbf{k})$, 
\begin{multline}
S_\alpha (\xi^* , \xi) = \sum_{j} \xi_{\alpha,\omega_j}^* G^{-1}_\alpha (\omega_j) \xi_{\alpha,\omega_j}^{\,}
\\ =
- \sum_{k,l} \xi_{\alpha,k}^* [-G^{-1}_{\alpha}]^{\,}_{kl} \xi_{\alpha,l}^{\,} .
\end{multline}
Integrating out the intermediate variables $\{\xi^*_{\alpha,k}\}_{k=1,\ldots ,M-1}$ and  $\{\xi^{\,}_{\alpha,l}\}_{l=2,\ldots, M}$ we have
\begin{equation}
\la  \xi_M | \hat{\rho} | \xi_1 \ra \propto 
e^{\sum_\alpha \xi^{*}_{\alpha,M} \xi^{\,}_{\alpha,1} 
\bigl\{ \bigl[ \frac{1}{M}\sum_j G_\alpha(\omega_j)e^{-i \omega_j} \bigr]^{-1} -1 \bigr\}  } .
\end{equation}
This implies that the density matrix only depends on the average occupations numbers $\bar{n}_\alpha$ of the states
\begin{equation}
\frac{1}{M}\sum_j G_\alpha(\omega_j)e^{-i \omega_j}  \equiv 1 - \bar{n}_\alpha .
\end{equation}
With the free fermion propagator of \eqref{eq:trialdensitymatrixparametrization} the frequency sum in the limit $M \rightarrow \infty$ gives the standard parametrization $\bar{n}_\alpha = (1+e^{\beta E_\alpha})^{-1}$, which also satisfy the necessary constraint that $0 \leq \bar{n}_\alpha \leq 1$. This implies that 
\begin{equation}
\la  \xi_M | \hat{\rho} | \xi_1 \ra \propto 
e^{\sum_\alpha \xi^{*}_{\alpha,M} \xi^{\,}_{\alpha,1} 
e^{-\beta E_{\alpha}}}
= \la \xi_M |e^{-\sum_\alpha \beta E_\alpha c^\dagger_\alpha c_\alpha } | \xi_1 \ra,
\end{equation}
where in the last step we have used the properties of fermion coherent states, see e.g. Ref.~\onlinecite{Robledo2009}.
Normalizing the density matrix properly it is now straightforward to calculate $S_{\hat{\rho}}$, and the result for the trial entropy becomes
\begin{multline}
S_{\hat{\rho}} = N \ln 2 - \sum_{a=0}^3 {\sum_{\mathbf{k}}}' 
\Bigl\{
\bar{n}_{\tilde{\gamma}_a} (\mathbf{k}) \ln [ \bar{n}_{\tilde{\gamma}_a} (\mathbf{k}) ]
\\ +
[1 - \bar{n}_{\tilde{\gamma}_a} (\mathbf{k}) ] \ln [1 - \bar{n}_{\tilde{\gamma}_a} (\mathbf{k}) ]
\Bigr\} .
\end{multline}
The first term is due to the spin system, which is maximally disordered. The second term is the usual entropy term for partially occupied fermion states. Extremizing the trial free energy we find that the trial energies should be chosen such that
\begin{equation}
E_{\alpha} = \frac{\delta  \text{Tr} \hat{\rho} \hat{H}}{\delta \bar{n}_{\alpha}},
\end{equation}
which are the standard mean field equations at finite temperature.\cite{FetterWalecka} It remains to choose the optimal $\alpha_{\mathbf{k}}$, which proceeds as in section~\ref{sec:operatorvariation}.

\section{Conclusions and Outlook}
\label{sec:conclusions}

We have studied the Kondo lattice starting from a faithful formulation of the model in terms of Majorana fermions.\cite{nilsson2011a} This formulation suggests a novel way of looking at the Kondo lattice: for strong coupling a good trial state is obtained by having ``deconfined'' independent Majorana fermions, while for weak coupling it is energetically favorable for the Majoranas coming from the $f$-spins to form ``confined'' three-body bound states. The real situation is of course something in between, and a mix of the two pictures. In this paper we have investigated this scenario through a relatively simple variational calculation. We have also formulated a discrete imaginary time path integral formalism for Majorana fermions, and we think that the result \eqref{eq:GrassmannMajoranaAction} is both simple and beautiful. It will hopefully be useful for further investigations of this and other models. The investigation presented here can be extended in several ways. One way is to allow for non-trivial spin correlations in the trial state as discussed in section \ref{sec:variationspincorrelation}.
Another way is to try to extend the trial density matrix calculation to allow for a frequency-dependent self-energy. Finally, the Majorana formalism could be applied to study the Kondo lattice model also away from half-filling, which is a straightforward extension of this work.

We wish to thank the Swedish research council (Vetenskapsr{\aa}det) for funding.

\section*{References}

\bibliography{MajoranaRefsIOP}

\end{document}